# Generating Packet-Level Header Traces Using GNN-powered GAN

Zhen Xu

College of Computer Science and Technology, Zhejiang University, Hangzhou, Zhejiang, 310000, China

3210105397@zju.edu.cn

**Abstract.** This study presents a novel method combining Graph Neural Networks (GNNs) and Generative Adversarial Networks (GANs) for generating packet-level header traces. By incorporating word2vec embeddings, this work significantly mitigates the dimensionality curse often associated with traditional one-hot encoding, thereby enhancing the training effectiveness of the model. Experimental results demonstrate that word2vec encoding captures semantic relationships between field values more effectively than one-hot encoding, improving the accuracy and naturalness of the generated data. Additionally, the introduction of GNNs further boosts the discriminator's ability to distinguish between real and synthetic data, leading to more realistic and diverse generated samples. The findings not only provide a new theoretical approach for network traffic data generation but also offer practical insights into improving data synthesis quality through enhanced feature representation and model architecture. Future research could focus on optimizing the integration of GNNs and GANs, reducing computational costs, and validating the model's generalizability on larger datasets. Exploring other encoding methods and model structure improvements may also yield new possibilities for network data generation. This research advances the field of data synthesis, with potential applications in network security and traffic analysis.

## 1. Introduction

Packet-level header traces encompass the metadata captured from network traffic at the packet level, including source and destination Internet Protocol (IP) addresses, port numbers, protocol types, and other header information [1,2]. These traces are crucial for various network management tasks, including the development of network detection and measurement algorithms. Nevertheless, acquiring such traces is frequently impeded by privacy concerns and commercial restrictions.

An effective alternative to direct trace collection is the generation of synthetic traces. Among various techniques, Generative Adversarial Networks (GANs) and Spatio-Temporal Adversarial Networks (STANs) have emerged as promising approaches [3,4]. GANs consist of a generator and a discriminator: the generator aims to create realistic synthetic data to deceive the discriminator, while the discriminator strives to distinguish between real and synthetic data. This adversarial training process enhances the generators: output quality, with the discriminators ability to differentiate playing a critical role in guiding the generators optimization.

However, traditional GAN models have predominantly been applied within the computer vision domain, and their efficacy in generating tabular data, such as packet-level header traces, remains limited [5]. Tabular data often comprises multiple fields with diverse characteristics, including discrete variables (e.g., port numbers, IP addresses, protocols, Type of Service) and continuous variables (e.g., Time to Live, timestamps, packet lengths). These fields typically exhibit complex distributions and interdependencies. Source and destination IP addresses frequently adhere to specific port ranges. Furthermore, there may be correlations between timestamps and packet lengths. Effectively capturing these intricate inter-field relationships poses a significant challenge when employing GANs for generating packet-level header traces.

This paper examines the impact of different field representation methods on the efficacy of data generation. Experimental results indicate that word2vec embeddings significantly outperform one-hot encoding [6]. While one-hot encoding represents each field as a high-dimensional sparse vector, greatly increasing the dimensionality and complexity of the input space, word2vec embeddings map fields to a low-dimensional continuous vector space [7]. This not only preserves the semantic relationships between fields but also reduces dimensionality, thereby facilitating the model's ability to capture distributions and correlations more effectively. Consequently, word2vec embeddings enhance the GAN training process, resulting in more accurate and realistic data samples.

Additionally, this paper introduces a novel approach to GAN training by incorporating a Graph Neural Network (GNN) to extract deep features from the fields. Traditional GAN discriminators rely solely on raw field features to assess data authenticity [8]. In contrast, the proposed method enriches the discriminator by integrating deep features extracted by the GNN, thus enhancing its discriminative capability. This dual-feature learning approach allows the discriminator to capture both superficial distributions and deeper structural relationships within the data. As a result, the discriminators supervision of the generator becomes more comprehensive, improving the realism and diversity of the generated data. Furthermore, this approach mitigates the curse of dimensionality, preserving information in high-dimensional feature spaces and enhancing model performance on complex datasets.

The contributions of this paper are as follows:

1. This work evaluates the performance of packet-level header traces represented by one-hot encoding versus word2vec embeddings in GAN training, demonstrating the superior efficacy of word2vec embeddings.

2. This work proposes an innovative GAN-based method for generating packet-level header traces, which encourages the generator to produce synthetic data that closely aligns with the deep features of the original data.

## 2. Method

### 2.1 Overview

The architecture of the proposed GNN-GAN is demonstrated in Fig. 1. Noise is input to the generator, which produces synthetic data. This data is then processed by a GNN to extract features, which are concatenated with the synthetic data. The combined features are fed into the discriminator to evaluate authenticity, and feedback is used to update the generator.

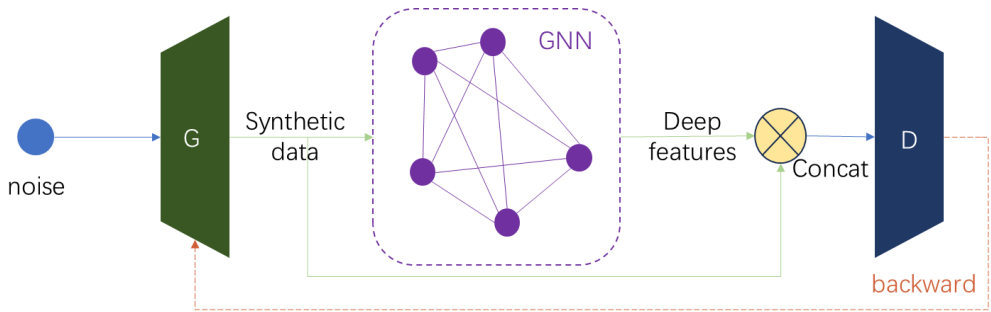

**Fig. 1** Overview of the GNN-GAN Framework (Figure Credits: Original).

### 2.2 One-Hot Encoding and Word2Vec Embedding

In synthetic data generation, the choice of field encoding methods significantly impacts model performance. This section discusses two common encoding techniques: One-Hot Encoding and Word2Vec Embedding.

### 2.2.1 One-Hot Encoding

One-Hot Encoding represents discrete features as high-dimensional sparse vectors, where each unique value is mapped to a binary vector with only one element set to 1. For example, if a field has three possible values (A, B, C), its One-Hot Encoding would be: A: [1, 0, 0], B: [0, 1, 0], C: [0, 0, 1].

This paper applied One-Hot Encoding to header traces. The specific methods are as follows. For discrete variables, such as IP address, port number, protocol type, Type of Service (TOS), each possible value is mapped to a unique index. For example, if the IP address field has K different values, the One-Hot encoded vector's dimension is K. For a specific IP address $v_i$, its One-Hot encoded vector is $o_i \in \{0,1\}^K$, where the $i$-th element is 1 and the rest are 0.

For continuous variables (such as TTL, packet length, timestamp), the continuous values are first divided into several buckets. Let the continuous value x be mapped to the $j$-th bucket out of $m$ buckets. The bucket index $j$ is then One-Hot encoded into an m-dimensional vector $b_j$. For example, if there are m buckets, the One-Hot Encoding of bucket index $j$ is $b_j \in \{0,1\}^m$, where the $j$-th element is 1 and the rest are 0.

In the encoding process, the original dimension of discrete fields increases from 1 to $K$, while the dimension of continuous fields increases from 1 to m (the number of buckets). Thus, the total encoding dimension is the sum of the encoded dimensions of discrete and continuous fields:

$$TotalDimensions = \sum_{i=1}^{n_d} K_i + \sum_{j=1}^{n_c} m_j \tag{1}$$

where $n_d$ is the number of discrete fields, $K_i$ is the number of possible values for the $i$-th discrete field, $n_c$ is the number of continuous fields, and $m_j$ is the number of buckets for the $j$-th continuous field.

**2.2.2 Word2Vec Embedding**

Word2Vec Embedding maps discrete features into a low-dimensional continuous vector space, providing richer and more meaningful feature representations. The key to this technique is its ability to capture semantic relationships between field values, so that similar features are mapped closer together in the vector space. For example, Word2Vec can map a specific port number like "port 80" to a dense low-dimensional vector, placing it near other port numbers with similar functions or meanings (such as "port 443") in the vector space, effectively reflecting the semantic similarity of these features.

This paper encoded the fields of header traces using Word2Vec. The specific steps are as follows:

Each discrete field in the header traces (such as IP addresses, port numbers, protocol types, etc.) is treated as an independent token. For example, the source IP address "192.168.1.1" is mapped to a low-dimensional vector $e_{scr\_IP}$, and the destination port 80 is mapped to a low-dimensional vector $e_{dst\_port}$. Using Word2Vec, each field's token is converted into an m-dimensional vector $e_i$, where $e_i \in R^m$ represents the embedding vector of that field.

The embedding vectors of each field are concatenated to form the feature representation of the entire header trace. For example, if a header trace contains $n$ fields (such as source IP, destination IP, source port, destination port, etc.), these fields' embedding vectors are concatenated in order, resulting in a combined vector **H** of dimension $n \times m$, where each $e_i$ represents the embedding of the corresponding field.

$$\mathbf{H} = [e_i, e_i, \dots, e_i] \in R^{n \times m} \tag{2}$$

This method not only effectively reduces the data's dimensionality but also preserves the semantic relationships between fields, providing stronger feature support for subsequent data analysis and generation tasks. By using Word2Vec embeddings, this work can capture the deep features of field values, enabling the generative model to better learn and generate synthetic data that aligns with the real data distribution.

Compared to One-Hot Encoding, the main advantages of Word2Vec Embedding include: (1) Dimensionality Reduction: High-dimensional sparse vectors are mapped to low-dimensional dense vectors, alleviating the curse of dimensionality. (2) Capturing Semantic Relationships: Able to capture deep relationships between field values, providing richer feature representations. (3)

Computational Efficiency: Low-dimensional vectors accelerate computation, making them suitable for deep learning models on large-scale data.

**2.3 Generative Adversarial Networks**

GANs are a type of deep learning model composed of a generator and a discriminator. The generator's goal is to generate data that is as realistic as possible to deceive the discriminator, while the discriminator is responsible for distinguishing between generated and real data. During training, the generator and discriminator compete with each other, with the generator trying to minimize the discriminator's ability to recognize fake data and the discriminator trying to maximize its accuracy in identifying real data.

The training of GANs can be described using the following mathematical formulation. Let G be the generator, D be the discriminator, x be the real data, and z be the noise input to the generator. The discriminator's goal is to maximize the probability of correctly classifying real and generated data, while the generator's goal is to minimize the probability of generated data being identified as fake by the discriminator. The objective function of GANs can be expressed as:

$$\min_G \max_D V(D,G) = E_{x \sim P_{data}(x)}[logD(x)] + E_{z \sim P_z(z)}[\log(1 - D(G(z)))] \qquad (3)$$

, where $E_{x \sim P_{data}(x)}[logD(x)]$ represents the expected value of the discriminator correctly classifying real data. $E_{z \sim P_z(z)}[\log(1 - D(G(x)))]$ represents the expected value of the discriminator classifying generated data as fake.

The generator's goal is to improve the data it generates by minimizing the following loss function:

$$\min_G -E_{z \sim P_z(z)}[\log(D(G(z)))] \qquad (4)$$

By optimizing these two objective functions, the generator and discriminator of GANs continuously improve, and eventually, the generator can generate synthetic data that is nearly indistinguishable from real data.

**2.4 Graph Neural Networks**

GNNs are a class of deep learning models designed to handle graph-structured data, aiming to capture the relationships between nodes and their neighbors within a graph. By propagating information across nodes in the graph, GNNs can effectively extract and learn deep features from graph data.

The fundamental idea of GNNs is to aggregate features of nodes based on the graph's adjacency structure. Specifically, the features of each node depend not only on itself but also on the features of its neighboring nodes. The main processes in GNNs include message passing, feature updating, and global aggregation.

Mathematically, the core operations of a GNN can be represented by the following steps.

First, message passing:

$$m_i^{(k)} = \sum_{j \in N(i)} W^{(k)} h_j^{(k-1)} \qquad (5)$$

, where $m_i^{(k)}$ denotes the message for node $i$ at the $k$-th layer, $N(i)$ represents the neighbors of node $i$, $W^{(k)}$ is the weight matrix at the $k$-th layer, and $h_j^{(k-1)}$ is the feature vector of the neighbor node $j$.

Second, feature updating:

$$h_j^{(k-1)} = Update(h_j^{(k-1)}, m_i^{(k)}) \qquad (6)$$

The feature updating operation combines the received messages with the node's own features to obtain the updated node features.

Third, global aggregation

$$h_G^{(k)} = Aggregate(\{h_j^{(K)}, |i \in V\}) \tag{7}$$

At the global level of the graph, the features of all nodes are aggregated to obtain a global representation of the graph.

**2.5 Integrating GNN with GAN**

This paper proposes a method for embedding GNNs into GANs to enhance the discriminator's discriminative capability. Specifically, the method involves first feeding real and synthetic data into the GNN to obtain deep feature representations. Let the input data be $X$ (where $X \in R^n$ represents the raw features of the data) and let $F_{GNN}$ (where $F_{GNN} \in R^m$) represent the embedded features obtained from the GNN. This process, as depicted in Figure 1, can be represented as:

$$F_{GNN} = GNN(X) \tag{8}$$

Next, the deep features $F_{GNN}$ extracted by the GNN are concatenated with the original data $X$. Let the concatenated features be $F_{concat}$, with dimensions (n+m), given by:

$$F_{concat} = [X, F_{GNN}] \in R^{(n+m)} \tag{9}$$

Then, the concatenated features $F_{concat}$ are input into the discriminator D for classification, and the discriminator's output represents the authenticity of the input data:

$$Output_D = D(F_{concat}) \tag{10}$$

For real data $X_{real}$ and synthetic data $X_{fake}$, they are processed through the GNN to obtain deep features $F_{GNN,real}$ and $F_{GNN,fake}$ respectively. The concatenation process can be represented as:

$$F_{concat,real} = [F_{real}, F_{GNN,real}] \in R^{(n+m)} \tag{11}$$
$$F_{concat,fake} = [F_{fake}, F_{GNN,fake}] \in R^{(n+m)} \tag{12}$$

Finally, the outputs of the discriminator for real and synthetic data are:

$$Output_{D,real} = D(F_{concat,real}) \tag{13}$$
$$Output_{D,fake} = D(F_{concat,fake}) \tag{14}$$

To further improve the GNN's feature extraction capability, this work incorporates an autoencoder framework where the GNN functions as an encoder. The goal is to train the GNN by compressing the input data $X$ into a lower-dimensional vector $Z$, which is then passed through a decoder to reconstruct the original input. This autoencoder setup helps in capturing the essential features of the data while ensuring that the GNN learns to maintain the integrity of the original data:

$$Z = Encoder(X) = GNN(X) \tag{15}$$
$$X' = Decoder(Z) \tag{16}$$

The objective is to minimize the reconstruction loss, ensuring that the decoded output $X'$ is as close as possible to the original input $X'$, i.e., $f(X) = X$. This reconstruction process refines the GNN's ability to generate meaningful embeddings, which in turn enhances the discriminator's performance in distinguishing real data from synthetic data.

With this approach, the discriminator learns to assess the authenticity of data from both the original data and the deep features extracted by the GNN, thereby enhancing the quality and diversity of the data generated by the generator.

## 3. Experiment and Result

### 3.1 Datasets

This work utilized the Center for Applied Internet Data Analysis (CAIDA) dataset, a widely recognized resource in the field of network research. The CAIDA dataset offers comprehensive and high-quality data collected from real-world internet traffic. This dataset is essential for evaluating network performance, security, and topology. Below is a brief overview of the CAIDA dataset: (1) Type of data: The CAIDA dataset includes various types of network data, such as anonymized internet traffic traces, internet topology information, and flow statistics. (2) Anonymized traffic traces: This part of the dataset contains anonymized network traffic captured over specified periods. All sensitive information, including IP addresses, is anonymized to ensure privacy while retaining the structure and flow of the traffic. (3) Internet topology data: The dataset provides detailed information about the internet's topology, including the relationships between different autonomous systems (AS) and their hierarchical structure. (4) Flow statistics: Summarized statistical data on network traffic is also available, offering insights into traffic patterns, bandwidth usage, and network behavior. (5) Usage: Researchers and engineers use CAIDA data for a range of applications, including network performance analysis, security monitoring, and the study of internet topology. The data helps in modeling network traffic, optimizing protocols, and understanding internet dynamics [9].

### 3.2 Results

To evaluate the effectiveness of different encoding and modeling strategies for generating packet header traces, the experiments compared three models: One-hot + Wasserstein-GAN (WGAN), Word2vec + WGAN, and Word2vec + GNN + WGAN [10]. The comparison was conducted using Jensen-Shannon (JS) Divergence for discrete variables and normalized Earth Mover's Distance (EMD) for continuous variables, both of which are crucial in assessing the fidelity of the generated data against real-world datasets.

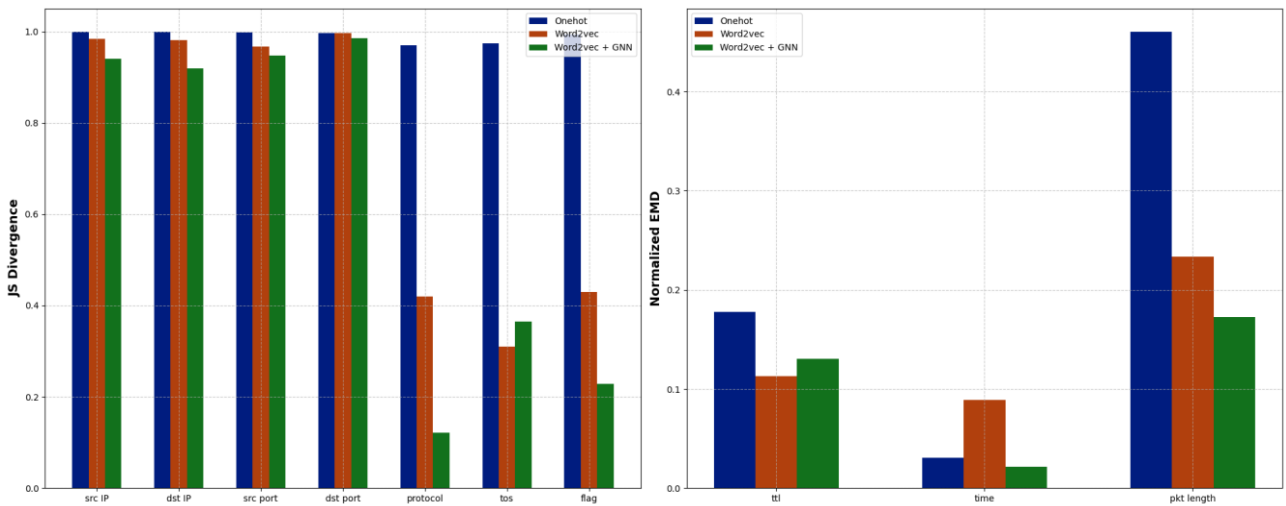

**Fig. 2** Comparison of JS Divergence (left) and normalized EMD (right) across different models and features (Figure Credits: Original).

Fig. 2 presents the JS Divergence (left) and normalized EMD (right) for the three models across various features. The left subplot demonstrates the JS Divergence for discrete variables including source IP, destination IP, source port, destination port, protocol, type id service, and flag. The right subplot displays the normalized EMD values for continuous variables such as time to live, time, and packet length. Each bar represents the performance of one of the three models, allowing for a direct comparison within each feature.

There are two key observations. (1) Word2vec Embedding Superiority: Across both JS Divergence and normalized EMD metrics, the Word2vec embedding approach consistently outperforms the traditional One-hot encoding, indicating that Word2vec captures the semantic relationships between features more effectively. (2) GNN Integration Enhances GAN Performance: The incorporation of GNN into the GAN framework (Word2vec + GNN + WGAN) further improves the model's ability to generate realistic header traces, as evidenced by the lower JS Divergence and normalized EMD scores. This suggests that leveraging GNN allows for better modeling of the underlying network structure, leading to more accurate generation of network traffic features.

These results underscore the importance of selecting appropriate feature representations and model architectures for generating high-fidelity synthetic network data. The Word2vec + GNN + WGAN model shows the most promise, particularly in scenarios where the precise replication of network traffic behavior is critical.

## 4. Discussion

This study introduces a novel method based on GNNs and GANs and validates its effectiveness in generating packet-level header traces. The experimental results show that, compared to traditional one-hot encoding, using word2vec embeddings significantly improves the quality of the generated data. This improvement is mainly due to word2vec's ability to better capture the semantic relationships between field values and reduce the impact of dimensionality explosion, enabling the generator to learn the true distribution of the data more effectively. Furthermore, incorporating GNNs enhances the discriminator's ability by extracting deep features, thereby increasing the authenticity and diversity of the generated data.

Despite the progress made in generating network traffic data, there are still some limitations. Firstly, the current model may be overfit to the specific features of certain datasets, limiting its applicability to different datasets. Secondly, the integration of GNNs and GANs increases the computational complexity of the model, which may become a bottleneck with large-scale datasets. Future research could focus on optimizing the model structure to reduce computational overhead and validating the model's generalizability on a broader range of datasets. Additionally, exploring other types of features encoding methods and improving the integration of GNNs and GANs are promising directions for further research. This will help to enhance the quality of generated data and expand the applicability of the method.

## 5. Conclusion

This study presents a novel approach that combines GNNs with GANs for generating packet-level header traces. By incorporating word2vec encoding, this work significantly reduces the dimensionality issues of GANs, enhancing the model's training effectiveness. Experimental results demonstrate that, compared to traditional one-hot encoding, word2vec encoding not only better captures the semantic relationships between field values but also improves the accuracy and naturalness of the generated data. Additionally, the inclusion of GNNs further enhances the discriminator's ability, allowing the model to generate synthetic samples that more closely resemble real data.

The results of this study provide a new theoretical perspective on network traffic data generation and practically demonstrate the potential for improving data synthesis quality through enhanced feature representation and model structure. This method can be applied in areas such as network security and traffic analysis, aiding in the development of advanced data synthesis tools and offering new technological pathways for addressing data privacy issues.

Future research could focus on further optimizing the integration of GNNs and GANs, reducing computational overhead, and validating its applicability on larger-scale datasets. Additionally, exploring other encoding methods and improvements in model structures will bring new possibilities to network data generation. It is believed that with these advancements, network data synthesis technology will experience further breakthroughs and innovations.

## References

[1] Dainotti Alberto, Antonio Pescapé, Giorgio Ventre. A packet-level characterization of network traffic. International Workshop on Computer-Aided Modeling, Analysis and Design of Communication Links and Networks, 2006: 38-45.

[2] Kim Seong Soo, AL Narasimha Reddy. Statistical techniques for detecting traffic anomalies through packet header data. IEEE/ACM Transactions on Networking, 2008, 16(3): 562-575.

[3] Goodfellow Ian, Pouget-Abadie Jean, Mirza Mehdi, et al. Generative adversarial networks. Communications of the ACM, 2020, 63(11): 139-144.

[4] Gao Nan, Xue Hao, Shao Wei, et al. Generative adversarial networks for spatio-temporal data: A survey. ACM Transactions on Intelligent Systems and Technology, 2022, 13(2): 1-25.

[5] Esfahani Shirin Nasr, Shahram Latifi. A survey of state-of-the-Art GAN-based approaches to image synthesis. Computer Science Conference Proceedings. 2019: 63-76.

[6] Mikolov Tomas, Chen Kai, Corrado Greg, et al. Efficient estimation of word representations in vector space. arXiv preprint, 2013: 1301.3781.

[7] Karani Dhruvil. Introduction to word embedding and word2vec. Towards Data Science, 2018, 1: 1-5.

[8] Wu Zonghan, Pan Shirui, Chen Fengwen, et al. A comprehensive survey on graph neural networks. IEEE transactions on neural networks and learning systems, 2020, 32(1): 4-24.

[9] CAIDA Data - Overview of Datasets, Monitors, and Reports. URL: https://www.caida.org/catalog/datasets/overview/. Last Accessed: 2024/08/11

[10] Martin Arjovsky, Soumith Chintala, Léon Bottou. Wasserstein GAN. arXiv preprint, 2017: 1701.07875.